# Microscopic signatures of Chern number sign reversal in twisted bilayer $WSe_2$

Ke Lv[1,2], Ya-Ning Ren[1,2,*], and Lin He[1,2,*]

**Affiliations:**
[1]Center for Advanced Quantum Studies, School of Physics and Astronomy, Beijing Normal University, Beijing 100875, China
[2]Key Laboratory of Multiscale Spin Physics, Ministry of Education, Beijing 100875, China

*Correspondence and requests for materials should be addressed to Ya-Ning Ren (e-mail: yning@mail.bnu.edu.cn), and Lin He (e-mail: helin@bnu.edu.cn).

**The discovery of quantized Chern numbers in twisted transition metal dichalcogenide (TMD) homobilayers—including 3.7° twisted $MoTe_2$ and 1.23° twisted $WSe_2$—has emerged as a defining breakthrough in physics[1-5]. A striking and unresolved puzzle from these studies is the unexpected opposite sign of the observed Chern numbers between the two systems. Recent theory has proposed a twist-angle-dependent Chern number sign reversal in both twisted $MoTe_2$ and $WSe_2$, offering a potential explanation for the disparate experimental observations[6]. However, a direct experimental verification of the twist-angle-dependent Chern number sign reversal in a specific twisted TMD homobilayer is still elusive. Here, we report the first experimental demonstration that the Chern numbers of the moiré frontier bands undergo sign reversal at a critical twist angle ~1.42° in twisted $WSe_2$ bilayers ($tWSe_2$). Using scanning tunnelling microscopy and spectroscopy, we direct measure layer-pseudospin skyrmion textures of $tWSe_2$ and our results reveal that $tWSe_2$ in the vicinity of the 1.42° exhibits a twist-angle-dependent layer-pseudospin polarization, an effect that serves as the fundamental origin of the observed Chern number sigh reversal[6-12].**

Moiré bilayers composed of semiconducting transition-metal dichalcogenides (TMDs) have emerged as exceptionally fertile platforms in recent years[1-26], attracting considerable interest due to their tunability for investigating novel strongly correlated phases of matter and quantum phenomena that arise from the interplay of correlation and topology. Notably, the topology of moiré bands in TMDs moiré lattices exhibits a strong dependence on the specific TMD species and the precise twist angle $\theta$ utilized to engineer the moiré superlattice. For instance, at a doping of one hole per moiré unit cell, recent experiments have yielded conflicting observations[1-5]: optical and transport characterization measurements have detected opposite Chern numbers in 3.7° twisted bilayer $MoTe_2$ (t$MoTe_2$) and 1.23° twisted bilayer $WSe_2$ (t$WSe_2$). Theoretically[6], by incorporating the competitive interplay between ferroelectricity and piezoelectricity in moiré superlattices, it is demonstrated that the Chern numbers of the moiré frontier bands exhibit twist-angle-driven sign reversal in both the t$MoTe_2$ and t$WSe_2$. This theoretical framework provides a viable mechanism for rationalizing the experimentally observed Chern number sign reversal in the 3.7° t$MoTe_2$ and 1.23° t$WSe_2$. However, direct experimental confirmation of the twist-angle-dependent Chern number sign reversal in a single specific twisted TMD homobilayer—a prerequisite for validating the proposed mechanism and unlocking rational design of topological moiré phases—remains an elusive goal.

In this work, we provide the first experimental demonstration that the Chern numbers of the moiré frontier bands undergo sign reversal at a critical twist angle ~1.42° in t$WSe_2$. The tunability of topological properties in TMD homobilayer can be traced to electronic polarization textures within moiré supercells, which form a layer-pseudospin skyrmion lattice that gives rise to non-trivial band topology via geometric phases[6-10]; continuum-model studies corroborate that this skyrmion structure is central to determining band topology, and scanning tunnelling microscopy (STM) is an ideal tool for probing such skyrmion-like pseudospin due to its capability to characterize both atomic and local electronic properties[11,12]. Using STM, we directly visualize layer-pseudospin skyrmion textures within the moiré unit cell of t$WSe_2$ with different twist angles, and our findings reveal that t$WSe_2$ close to the 1.42° critical twist angle exhibit a twist-angle-dependent layer-pseudospin polarization, an effect that underpins the observed Chern number sign reversal[6].

**Physical pictures**

In twisted TMDs homobilayers, the twist angle emerges as a versatile and robust tuning knob that dictates the band topological landscape[1-5]. A hallmark of such twist-engineered topology is the topological transition, which manifests as a pronounced redistribution of electron wavefunctions across high-symmetry regions within the moiré unit cell—an effect driven by subtle variations in the strength of moiré potential[6-12]. The evolution of real-space wavefunction localization is directly encoded in the z-component of the layer pseudospin: its winding number not only encapsulates the intrinsic topological character of the bands but also forges an unambiguous link between real-space localization and band topology[6-12]. Figure 1a depicts the moiré supercell of twisted WSe$_2$ homobilayers, which is inherently partitioned into three distinct stacking domains: XM, in which a chalcogen atom sits atop a metal atom; MX, in which a metal atom sits atop a chalcogen atom; and MM, in which a metal atom sits atop a metal atom. As theoretically predicted[6], the flat-band wavefunction localizes in the XM regions of the top layer and the MX regions of the bottom layer when the twist angle smaller than the critical value $\theta_0$ (theoretically ~ 1.54°), as illustrated in Fig. 1b. This layer-dependent wavefunction configuration, protected by time-reversal symmetry T and $C_{2x}$ symmetry, gives rise to a winding number of −1, which translates to a Chern number of −1 for the topmost flat band, as dictated by the wavefunction's evolution across the moiré supercell. At $\theta_0$, the flat-band wavefunction exhibits symmetric localization in MX and XM regions. This symmetric distribution yields a uniform layer pseudospin across the moiré supercell, corresponding to a winding number of zero and thus a topologically trivial band (Fig. 1c). For twist angle exceeds $\theta_0$, the flat-band wavefunction shifts to the MX regions of the top layer and the XM regions of the bottom layer, reversing the pseudospin winding direction and flipping the Chern number of the topmost flat band (Fig. 1d). In the following, we directly probe this twist-angle-dependent topological transition via STM, by tracking the real-space localization of the flat-band wavefunctions, i.e., the z component of layer-pseudospin.

**Experimental observation**

Figure 2a illustrates the experimental setup. High-quality twisted WSe$_2$ homobilayers were fabricated atop monolayer graphene and hexagonal boron nitride (h-BN) via the conventional "tear-and-stack" technique[15,16,23,27-30] (see Methods and Extended Fig. 1 for details). Here, graphene served exclusively as a local drain electrode for tunnelling measurements. Figure 2b presents a representative STM topography of tWSe$_2$. The moiré period $D$ of the tWSe$_2$ exhibits slight variation from left to right, attributed to residual strain in the heterostructure. Local twist angles $\theta$ of the tWSe$_2$ were extracted using the relation $D = a/(2\sin(\theta/2))$, where $a$ = 0.328 nm denotes the lattice constant of WSe$_2$. This analysis reveals a continuous $\theta$ variation from ~1.19° (left) to ~1.48° (right). Corresponding zoomed-in topographic images of tWSe$_2$ at these twist angles (1.19°, 1.25°, and 1.48°) are presented in Figs. 2c-e, respectively. Here, we note that continuous twist-angle tunability enables unambiguous delineation of the XM and MX regions within the tWSe$_2$ moiré supercell across the full range of twist angles investigated, thereby facilitating systematic probing of their twist-angle-dependent electronic characteristic.

In tWSe$_2$, valence-band edge states, derived from the K point of the monolayer Brillouin zone, are relatively flat, isolated, and host non-zero valley Chern numbers, by contrast,

Γ-point states are energetically well-separated, first emerging ~200 meV below the valence-band edge (the energy difference between the K- and Γ-point states depends on the specific TMD species and on the twist angle)[6-10]. Notably, the lowest few K-point-derived valence-bands exhibit a large tunnelling decay constant, rendering them weakly detectable via conventional constant-height scanning tunnelling spectroscopy (STS), which preferentially probes Γ-valley states. Recent advances constant-current STS have enhanced sensitivity to K-valley states by reducing the tip-sample distance, thereby enabling tunnelling to these otherwise weakly coupled states[11,12,31,32]. Herein, we resolve these two valley-specific states individually using a hybrid strategy combining constant-height STS with constant-current STS. To compare electronic wavefunction distribution characteristics, we performed constant-current (K-state detection) and constant-height (Γ-state detection) STS measurements along the MM–XM–MX-MM direction of tWSe$_2$ with twist angles spanning 1.19° — 1.48°. As shown in Figs. 2f-k, the relative LDOS intensities of MM, MX, and XM stacking, along with their evolution upon decreasing twist angle, are clearly visualized in the angle-ascending sequence. The energy difference of the uppermost Γ bands between MM sites and MX/XM sites reflects a large moiré potential modulation amplitude of about 200 meV for the Γ valley (Figs. 2i-k). In contrast, the K-valley states, lying ~200 meV above Γ-valley states, exhibit a weaker peak position variation, ~50 meV (Figs. 2f-h). Both the distinct moiré potential modulation amplitudes for K- and Γ-valley states and the energy offset between K- and Γ-point states are consistent with previous calculations and experimental results[6-12]. Unlike Γ states (Figs. 2i-k), which show nearly uniform intensity across MX and XM regions, K-state intensity displays a striking contrast, especially between MX and XM sites. This contrast arises from opposite layer polarizations: for instance, for tWSe$_2$ at 1.25°, the dipole points downward at MX site and upward at XM site, resulting in opposite wavefunction localization in the top (XM) and bottom (MX) layers. The most striking feature is the abrupt inversion of the relative LDOS contrast between MX and XM sites when the twist angle crosses a critical threshold near 1.42° (see Extended Fig. 2 for experimental results of 1.42° tWSe$_2$). For instance, in tWSe$_2$ with a twist angle of 1.25°, K-state intensity at XM site exceeds that at the MX site; conversely, the opposite trend holds for tWSe$_2$ at 1.48°. This distinct inversion occurs exclusively in the K-state LDOS, being most pronounced at the uppermost STS peak and persisting over a broad energy window. Notably, this observation aligns with the predicted twist-angle-dependent Chern number sign reversal in twisted TMD homobilayer[6].

To further elucidate how the spatial distribution of the LDOS evolves across the moiré supercell as a function of twist angle, we acquired constant-current (K-state detection) and constant-height (Γ-state detection) STS maps for tWSe$_2$ with different twist angles, as summarized in Fig. 3 (see Extended Figs. 2-9 for more experimental results). For tWSe$_2$ with a twist angle of 1.19°, the moiré frontier flat-band states originating from the K points exhibit markedly stronger intensity in the XM region than in MX region (Fig. 3a). This trend is replicated for tWSe$_2$ at 1.25° (Fig. 3b). When the twist angle is tuned to the critical threshold of 1.42°, the LDOS contrast between MX and XM sites becomes nearly indistinguishable (see Extended Fig. 2 for data at this critical twist angle). By contrast, when the twist angle is increased beyond this critical twist angle, for instance, to 1.48° in tWSe$_2$, the LDOS intensity in the MX region is significantly enhanced, ultimately surpassing that in the XM region (Fig. 3c). This evolution of LDOS localization within the moiré supercell is consistent with the spatially resolved STS spectra presented in Fig. 2. Notably, no analogous transition is observed in the Γ-state STS maps (Figs. 3d-f), where Γ states display nearly uniform intensity across MX and XM regions in tWSe$_2$ regardless of twist angle (see Extended Figs. 3, 5, and 7 for more experimental results). This twist-angle-dependent inversion of K-valley LDOS intensity across high-symmetry stacking regions represents the first experimental confirmation of this phenomenon in twisted TMD homobilayers. Briefly, the LDOS inversion between XM and MX regions near the critical angle is attributed to the competition between piezoelectric and out-of-plane ferroelectric polarizations[6]. Above the critical angle, the intrinsic piezoelectric effect of the TMD monolayer dominates, leading to stronger electron wavefunction localization at the MX regions. At the critical angle, these two polarizations are precisely balanced. Below the critical twist angle, the more prominent relaxation effect of the moiré superlattice take precedence, driving polarization toward XM regions. Experimentally, a further reduction in twist angle (e.g., from 1.25° to 1.19°) yields a more pronounced LDOS contrast between XM and MX regions (Figs. 3a, 3b and Extended Figs. 4, 6, and 9). This observation stems from enhanced moiré relaxation effect in high-symmetry moiré regions, providing further experimental support for the theoretically proposed mechanism[6]. Collectively, these observations provide direct real-space evidence for the twist-angle-dependent relocation of the flat-band wavefunction, underscoring the intrinsic connection between the moiré superlattice geometry and the resulting topological band properties.

**Conclusions**

In summary, we report the first experimental realization of a twist-angle–dependent inversion of the flat-band Chern number in tWSe$_2$, achieved via direct real-space imaging of flat-band LDOS localization. By systematically characterizing a moiré superlattice region with a continuous, gradual twist-angle variation, we demonstrate continuous tunability of the band-topological properties, establishing layer pseudospin as a key quantum degree of freedom for characterizing electron wavefunctions in twisted TMD homobilayers. The microscopic observation reported in this work not only resolves the long-standing puzzle of conflicting Chern number signs in twisted TMDs but also establishes a new pathway for engineering topological phases via precise control of twist angle and pseudospin order.

## Methods

### Device fabrication

The device was fabricated by using the "tear and stack" method[15,16,23,27-30]. A h-BN substrate was picked up by Polydimethylsiloxane (PDMS) film while the graphene and monolayer $WSe_2$ are exfoliated on $SiO_2$/Si chips, the exfoliated monolayer $WSe_2$ was pre-cut into two parts by atomic force microscopy (AFM) tip. Next, the three parts were picked by the thick h-BN piece by piece with manually adjusted twist angles. Finally, with the help of another PDMS, the whole heterostructure was upturned and transferred to the $SiO_2$/Si chip with pre-coated Au/Cr electrode and the graphite flake was used to connect the electrode with the sample. The detailed procedure and photos of sample can be seen in Extended Fig. 1.

### STM measurements

STM/STS measurements were performed in low-temperature (78 K) and ultrahigh vacuum (~$10^{-10}$ Torr) scanning probe microscopes [USM-1400] from UNISOKU. The tips were obtained by chemical etching from a tungsten wire. The differential conductance (d$I$/d$V$) measurements were taken by a standard lock-in technique with an ac bias modulation of 2 mV and 793 Hz signal added to the tunnelling bias and performed in a constant-current mode, where the current feedback was left on while the bias voltage changed allowing the tip to change its height, whereas for constant height mode, the feedback is turned off during the STS measurements. Electrochemically etched tungsten tips were used for imaging and spectroscopy.

### Acknowledgements

This work was supported by the National Natural Science Foundation of China (Grant Nos. 12425405, 12404198, 12141401), National Key R and D Program of China (Grant Nos. 2021YFA1400100, 2021YFA1401900), "the Fundamental Research Funds for the Central Universities" (Grant No. 310400209521), the China National Postdoctoral Program for Innovative Talents (BX20240040), the China Postdoctoral Science Foundation (2023M740296). The devices were fabricated using the transfer platform from Shanghai Onway Technology Co., Ltd.

### Contributions

L.H. and Y.-N.R. conceived the work and designed the research strategy. K.L. fabricated the samples and performed the measurements. L.H., Y.-N.R. and K.L. analyzed the experimental data and wrote the paper together.

Corresponding authors

Correspondence to Ya-Ning Ren, Lin He.

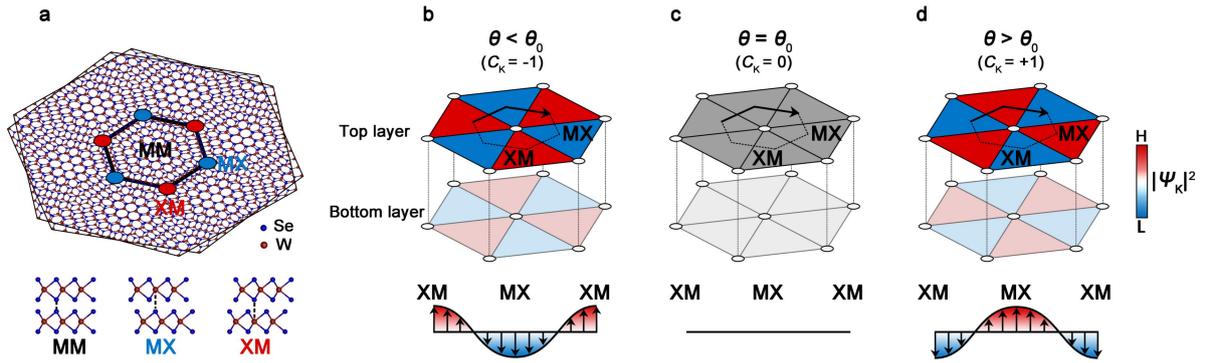

**Fig. 1 | Twist-angle-controlled layer-pseudospin polarization and Chern number in twisted bilayer TMDs. a**, Schematic illustration of the moiré superlattice in tWSe$_2$, comprising high-symmetry stacking regions (MM, MX and XM), with MX and XM sites forming a honeycomb arrangement. **b-d**, Top panels: schematics of the spatial localization $|\Psi_K|^2$ of the topmost valence flat-band wavefunctions at different twist angles. Bottom panels: evolution of the out-of-plane (z-component) layer pseudospin distribution along the direction indicated by the arrows in the top panels for twisted bilayer TMDs at different twist angles.

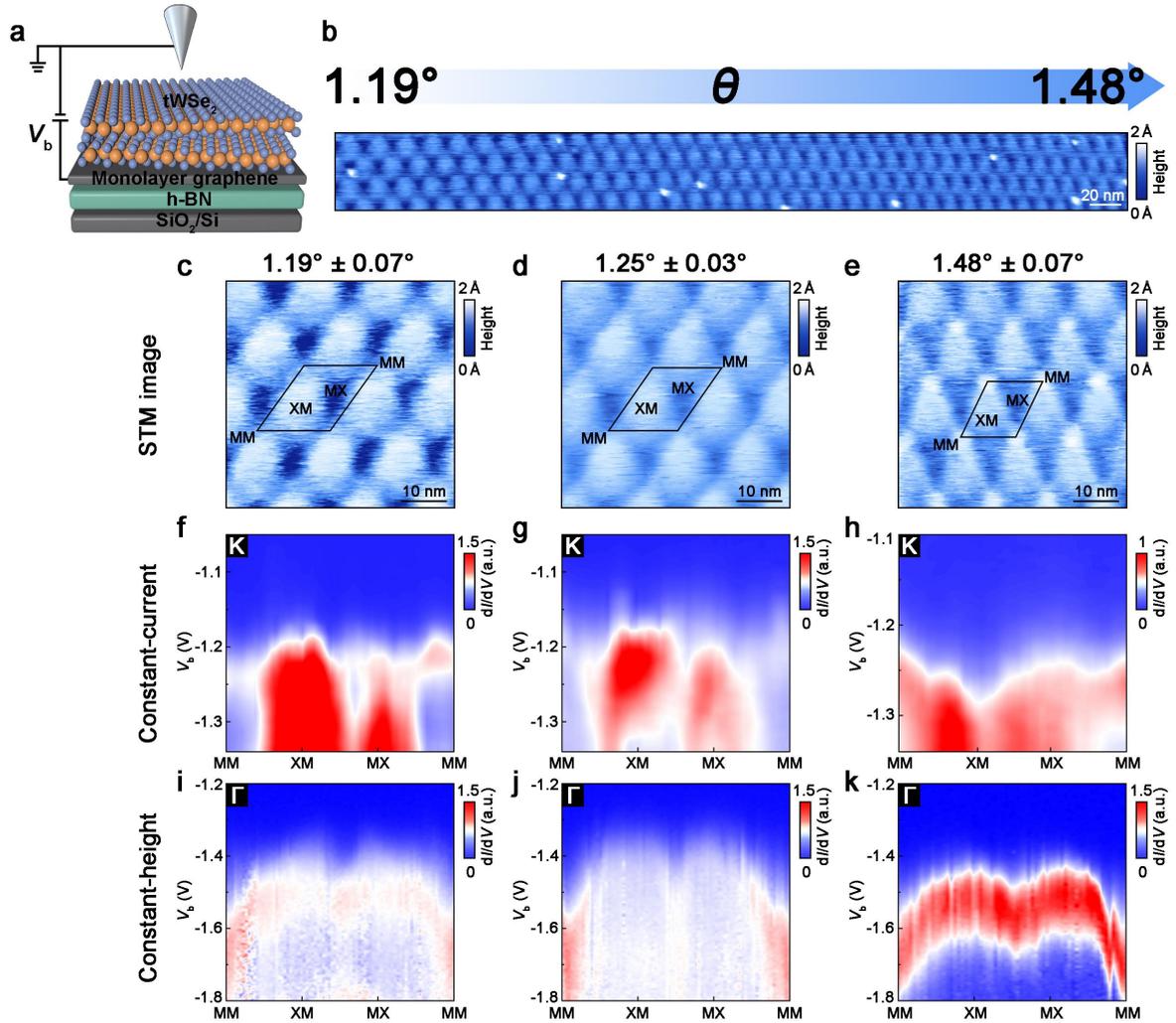

**Fig. 2 | STM and STS characterizations of tWSe$_2$ with varying twist angles. a**, Schematic illustration of the STM measurement setup for tWSe$_2$. **b**, STM image ($V_b$ = -1.80 V, $I$ = 100 pA) of tWSe$_2$ region exhibiting a continuous twist-angle variation (1.19° to 1.48°) induced by slight strain. **c-e**, Zoomed-in STM topographic images of tWSe$_2$ with distinct twist angles (corresponding to 1.19°, 1.25°, and 1.48°, respectively). Scanning parameters: (**c**) $V_b$ = -1.80 V, $I$ = 260 pA; (**d**) $V_b$ = –1.79 V, $I$ = 300 pA; (**e**) $V_b$ = –1.80 V, $I$ = 258 pA. **f-h**, Constant-current STS measurements showing the spatial distribution of K states along the MM-XM-MX-MM direction for tWSe$_2$ at different twist angles. **i-k**, Constant-height STS measurements revealing the spatial distribution of Γ states along the MM-XM-MX-MM direction for tWSe$_2$ at different twist angles.

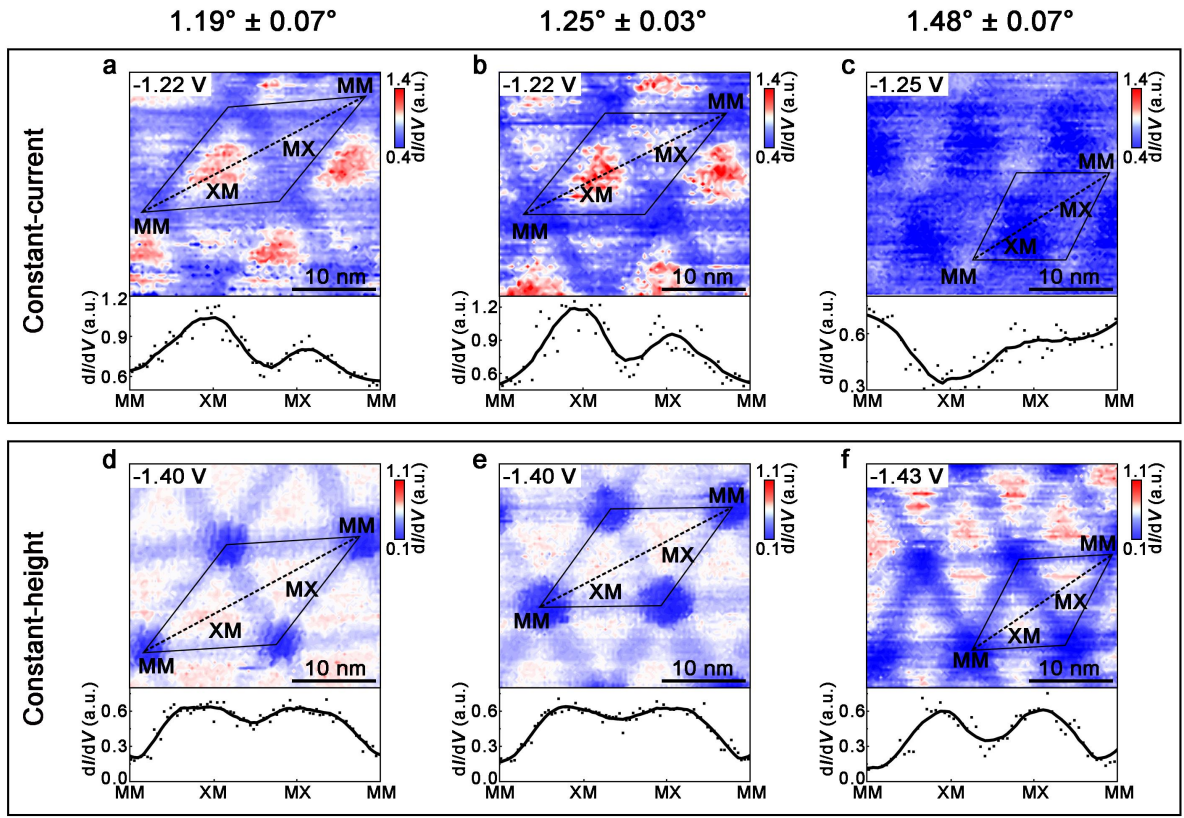

**Fig. 3 | Localization of K- and Γ-valley states in tWSe$_2$ with different twist angles. a-c**, Top panels: constant-current STS maps of the K-valley topmost valence flat band in tWSe$_2$ at different twist angles. The tunnelling currents *I* are 60 pA, 45 pA, and 50 pA for panels **a-c**, respectively. Bottom panels: the radial differential tunnelling conductance profiles taken along the black linecuts indicated in the corresponding top panels. The solid dots represent the experimental data, and the solid line denotes the smoothed curve. **d-f**, Top panels: constant-height STS maps of the Γ-valley topmost valence flat band in tWSe$_2$ at different twist angles. The initial tunnelling current *I* is 300 pA for all three panels. Bottom panel: the radial differential tunnelling conductance profiles taken along the black linecuts indicated in the corresponding top panels. The solid dots represent the experimental data, and the solid line denotes the smoothed curve.

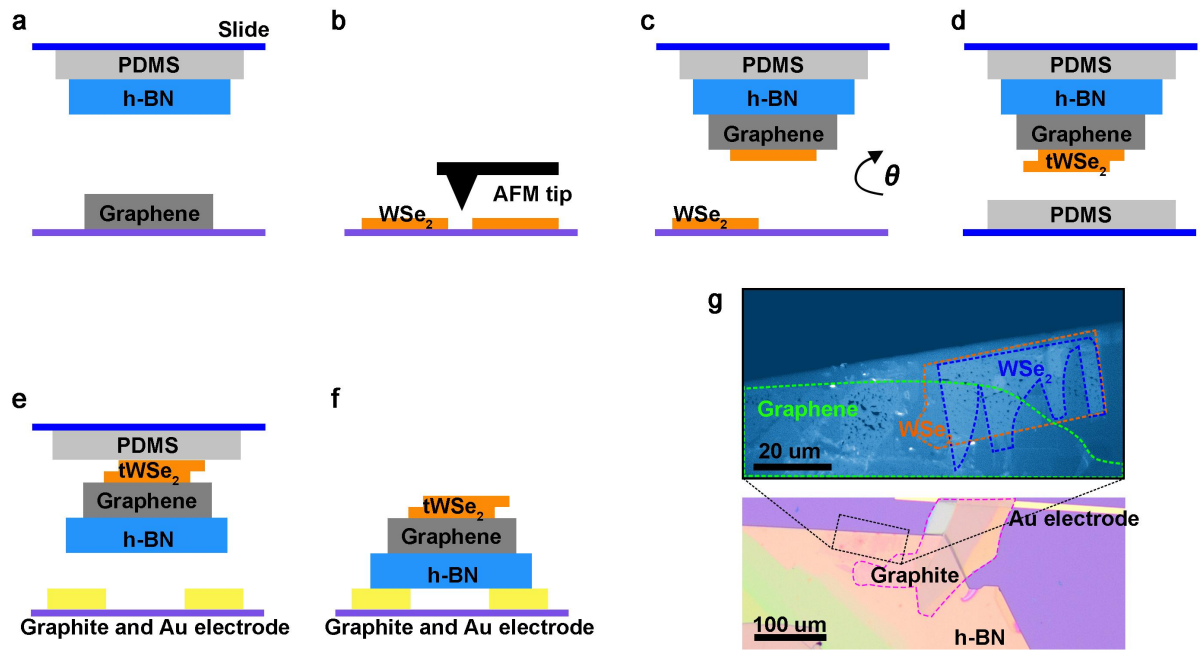

**Extended Data Fig. 1 | Fabrication of the tWSe$_2$ device. a-f**, Step-by-step schematic of the fabrication procedure used to assemble the tWSe$_2$ device (see Methods for detailed processes). **g**, Optical microscope image of the resulting tWSe$_2$ structure, where the blue and orange outlines mark the two individual WSe$_2$ monolayers assembled with a controlled relative twist angle $\theta$.

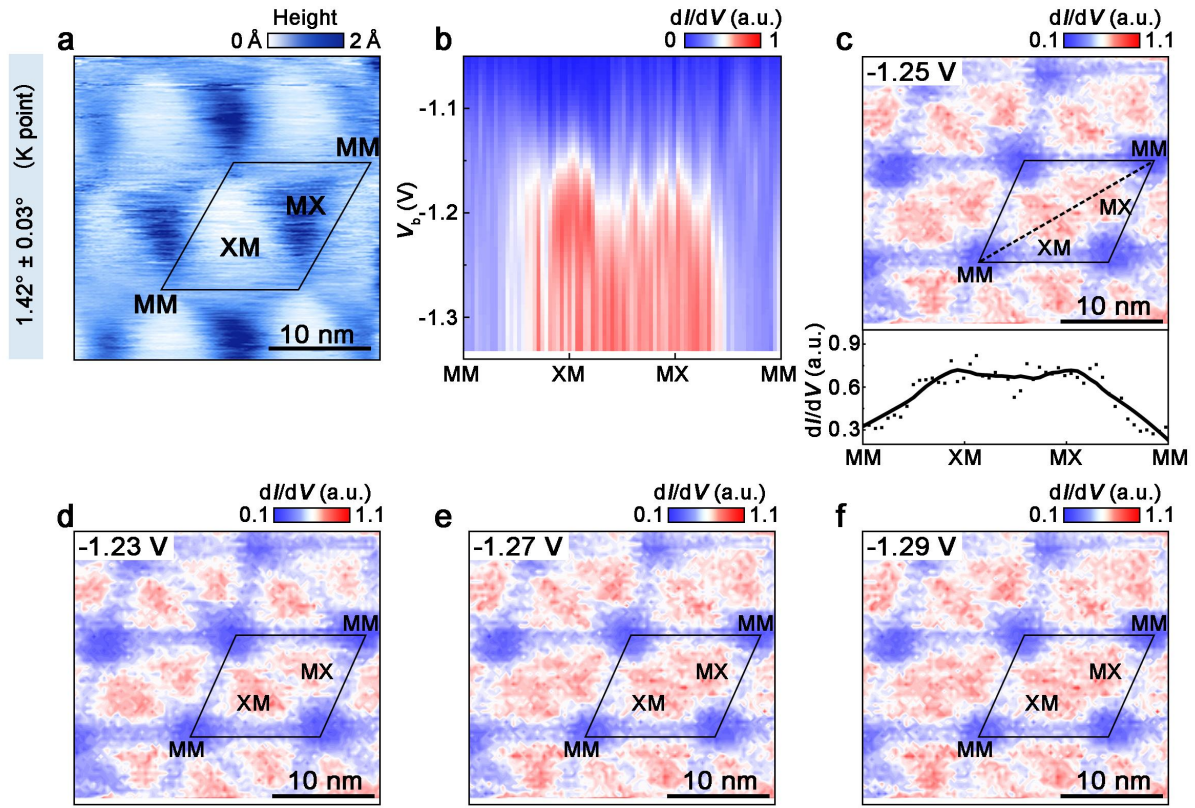

**Extended Data Fig. 2 | Localization of K-valley states at a twist angle of ~1.42°. a**, STM topographic images ($V_b$ = -1.75 V, $I$ = 210 pA) of tWSe$_2$ with a twist angle of 1.42°. **b**, Constant-current STS measurements showing the spatial distribution of K-valley states along the MM-XM-MX-MM direction. **c-f**, Differential tunnelling conductance maps measured at the indicated bias voltages. The initial tunnelling current is 30 pA. Bottom panel of **c**: radial differential tunnelling conductance profiles taken along the black linecut shown in the top panel of **c**. The solid dots represent the experimental data, and the solid line denotes the smoothed curve.

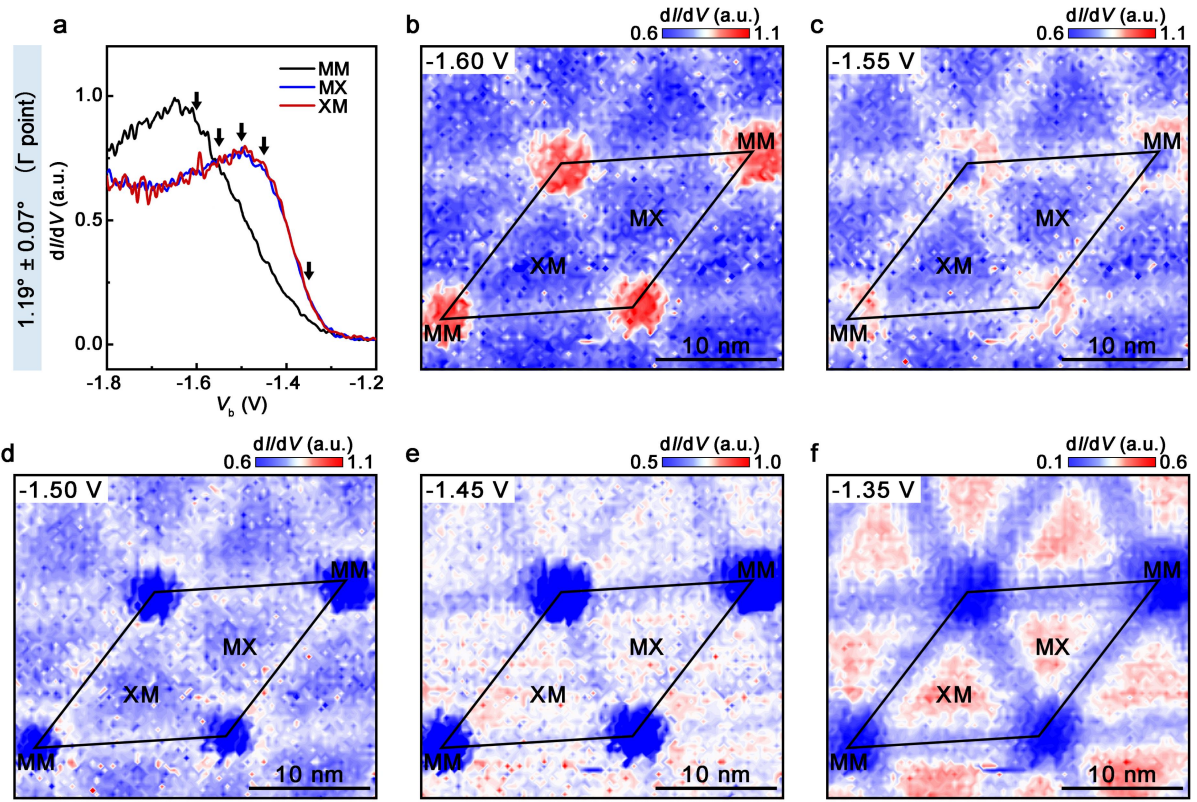

**Extended Data Fig. 3 | Localization of Γ-valley states at a twist angle of ~1.19°. a**, Representative constant-height d$I$/d$V$ spectra acquired at three high-symmetry sites—MM, MX, and XM. **b-f**, Differential tunnelling conductance maps measured at the bias voltages indicated by the arrows in **a**. The initial tunnelling current is 300 pA.

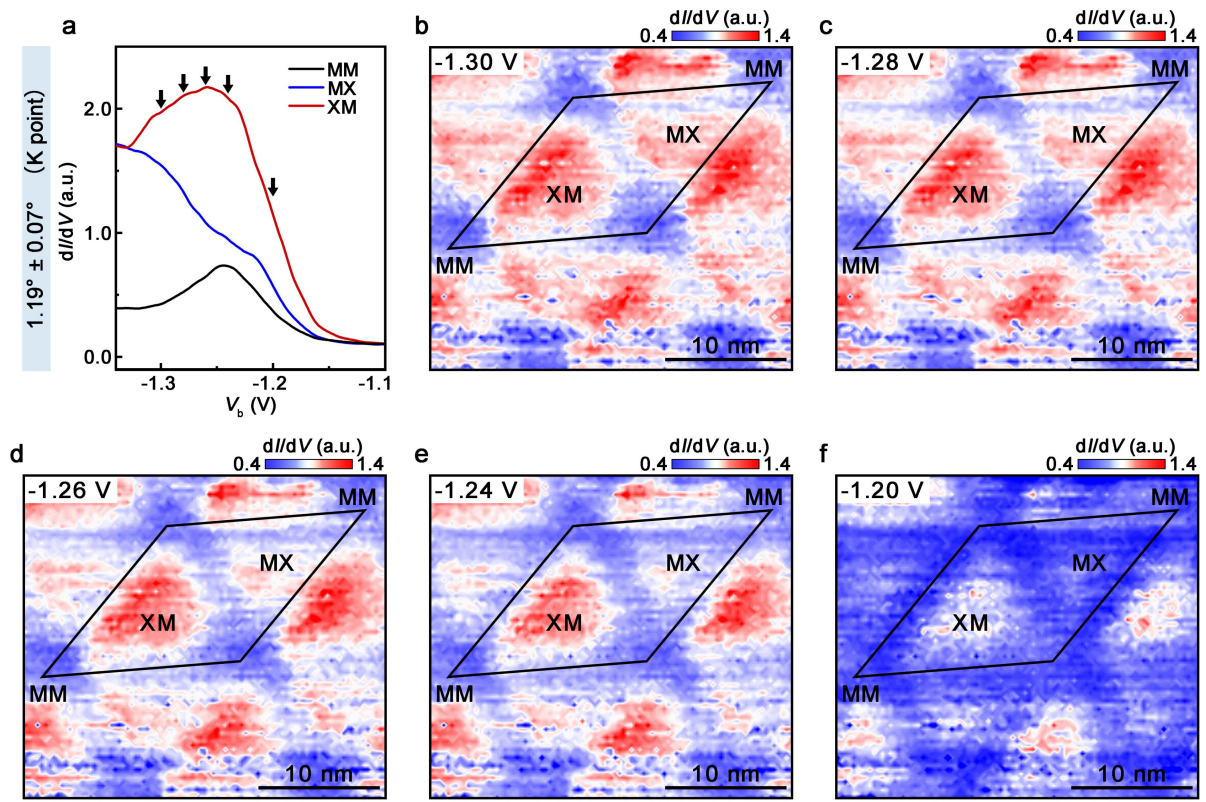

**Extended Data Fig. 4 | Localization of K-valley states at a twist angle of ~1.19°. a**, Representative constant-current d$I$/d$V$ spectra acquired at three high-symmetry sites—MM, MX, and XM. **b-f**, Differential tunnelling conductance maps measured at the bias voltages indicated by the arrows in **a**. The initial tunnelling current is 60 pA.

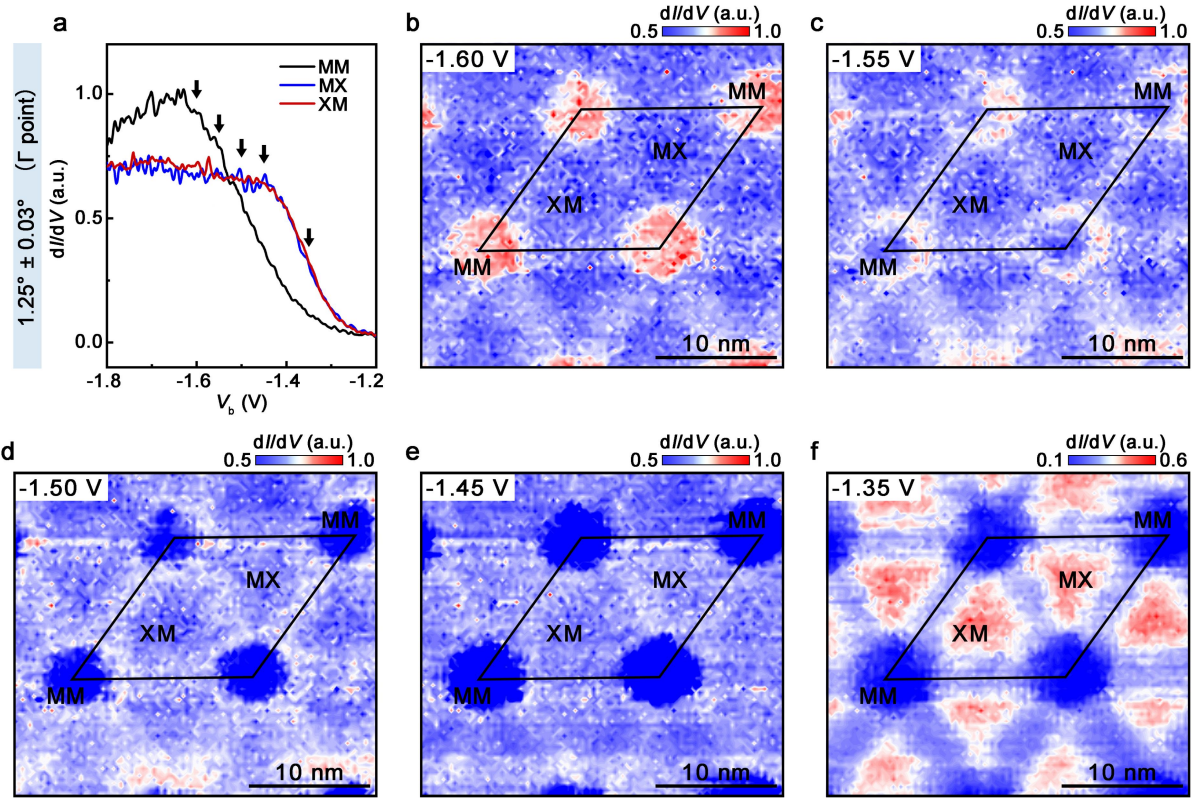

**Extended Data Fig. 5 | Localization of Γ-valley states at a twist angle of ~1.25°. a**, Representative constant-height d$I$/d$V$ spectra acquired at three high-symmetry sites—MM, MX, and XM. **b-f**, Differential tunnelling conductance maps measured at the bias voltages indicated by the arrows in **a**. The initial tunnelling current is 300 pA.

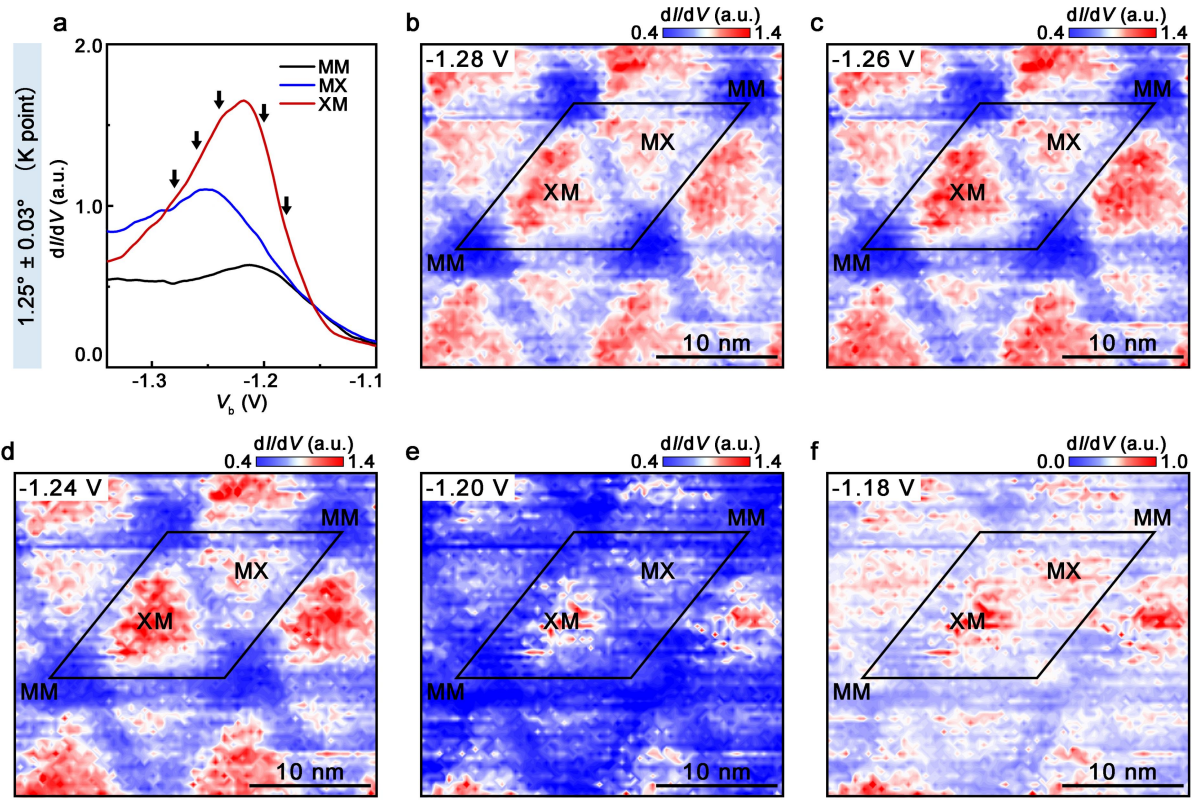

**Extended Data Fig. 6 | Localization of K-valley states at a twist angle of ~1.25°. a**, Representative constant-current d$I$/d$V$ spectra acquired at three high-symmetry sites—MM, MX, and XM. **b-f**, Differential tunnelling conductance maps measured at the bias voltages indicated by the arrows in **a**. The initial tunnelling current is 45 pA.

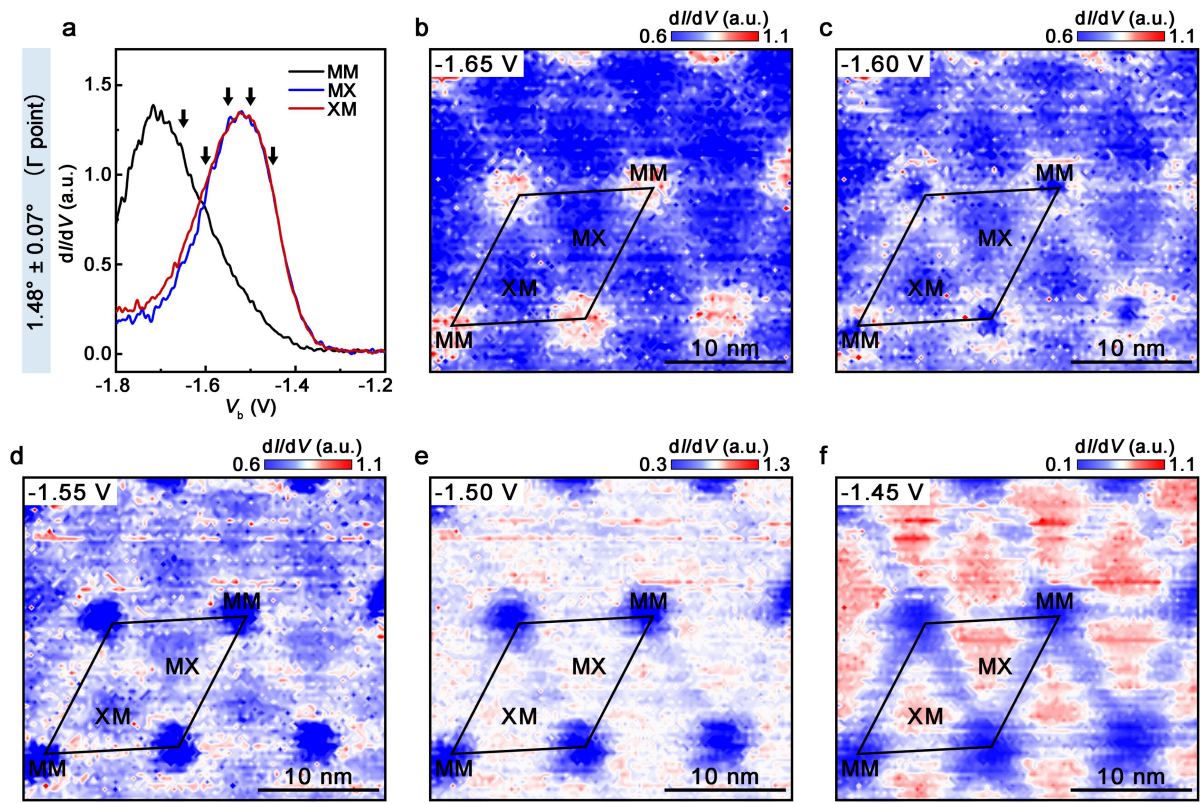

**Extended Data Fig. 7 | Localization of Γ-valley states at a twist angle of ~1.48°. a**, Representative constant-height d$I$/d$V$ spectra acquired at three high-symmetry sites—MM, MX, and XM. **b-f**, Differential tunnelling conductance maps measured at the bias voltages indicated by the arrows in **a**. The initial tunnelling current is 300 pA.

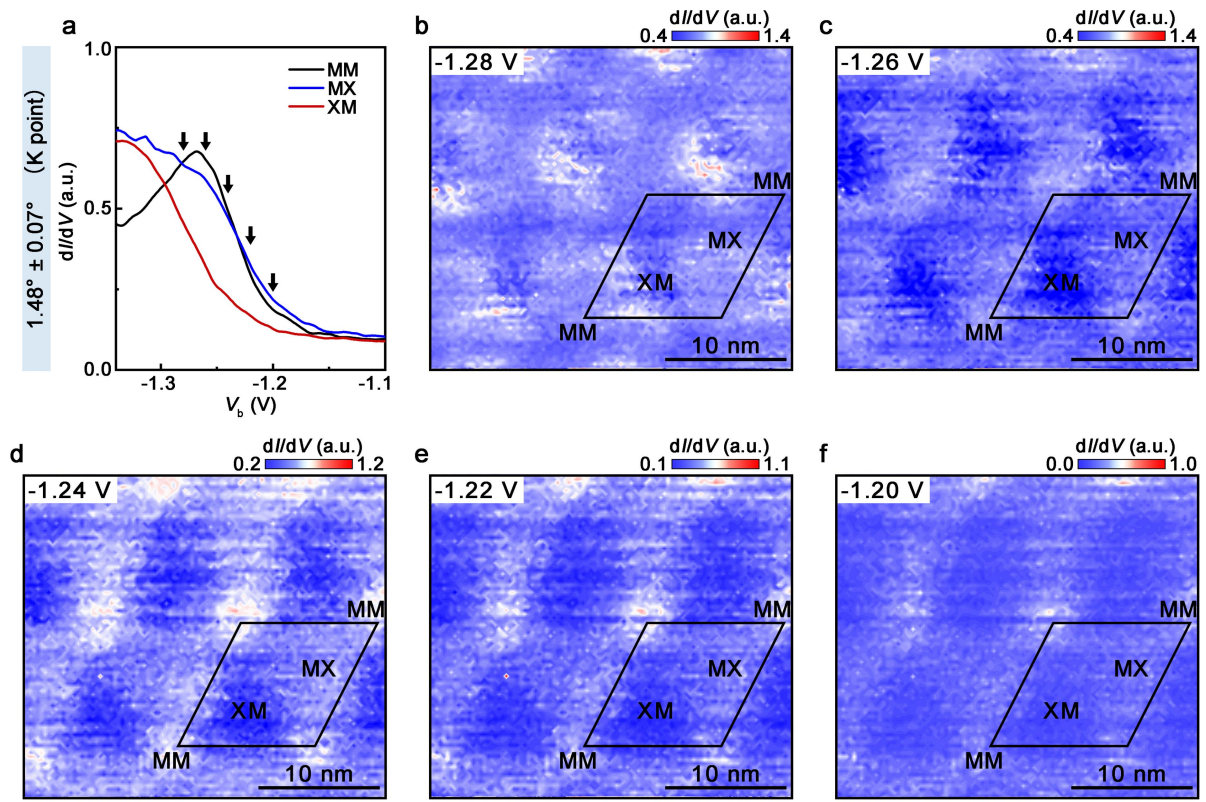

**Extended Data Fig. 8 | Localization of K-valley states at a twist angle of ~1.48°. a**, Representative constant-current d$I$/d$V$ spectra acquired at three high-symmetry sites—MM, MX, and XM. **b-f**, Differential tunnelling conductance maps measured at the bias voltages indicated by the arrows in **a**. The initial tunnelling current is 50 pA.

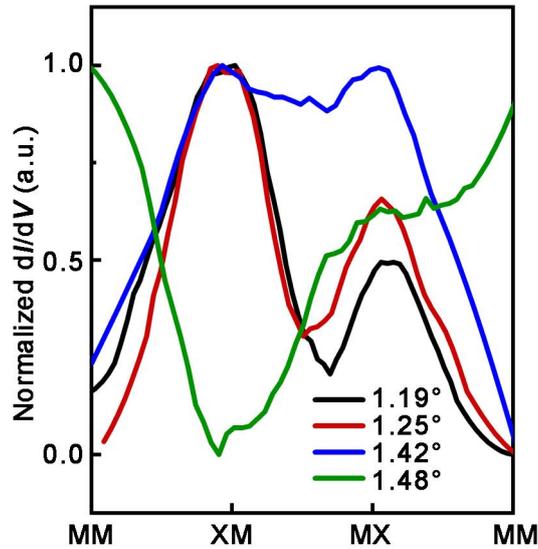

**Extended Data Fig. 9 | Twist-angle dependence of radial differential tunnelling conductance.** Summary of the normalized radial differential tunnelling conductance profiles measured for twist angles from 1.19° to 1.48°. By comparison, at small twist angles (1.19° and 1.25°), the XM region exhibits a higher density of states than the MX region. As the twist angle increases, the contrast between the XM and MX densities of states gradually decreases. Near the critical angle of ~1.45°, the two become nearly identical. For larger twist angles, the density-of-states contrast reverses, with the MX region exhibiting a higher density of states than the XM region.